\documentclass{INTERSPEECH2023}

\usepackage{multirow,adjustbox,amsmath}
\usepackage[table,x11names, svgnames, dvipsnames]{xcolor}

\interspeechcameraready 

\title{Adapter Incremental Continual Learning of Efficient Audio Spectrogram Transformers}

\title{Adapter Incremental Continual Learning of Efficient Audio Spectrogram Transformers}
\name{Nithish Muthuchamy Selvaraj*$^1$, Xiaobao Guo*$^{12}$, Adams Kong$^2$, Bingquan Shen$^3$, Alex Kot$^1$}
\address{
  $^1$Rapid-Rich Object Search (ROSE) Lab, Nanyang Technological University, Singapore\\
  $^2$School of Computer Science and Engineering, Nanyang Technological University, Singapore\\
  $^3$DSO National Laboratories, Singapore
  }
\email{
}

\begin{document}

\maketitle

\def\thefootnote{}\footnotetext{\textbf{Published as a conference paper in INTERSPEECH 2023.}}
\def\thefootnote{*}\footnotetext{These authors contributed equally to this work.}
 
\begin{abstract}

Efficient tuning of neural networks for continual learning with minimal computational resources remains a challenge. In this paper, we propose continual learning of audio classifiers with parameter and compute efficient Audio Spectrogram Transformers (AST). To reduce the trainable parameters without performance degradation we propose AST with Convolutional Adapter, which has less than 5\% of trainable parameters of full fine-tuning. To reduce the computational complexity of self-attention, we introduce a novel Frequency-Time factorized Attention (FTA) method that achieves competitive performance with only a factor of the computations. Finally, we formulate our method called Adapter Incremental Continual Learning (AI-CL), as a combination of the parameter-efficient Convolutional Adapter and the compute-efficient FTA. Experiments on ESC-50, SpeechCommandsV2, and Audio-Visual Event benchmarks show that our proposed method efficiently learns new tasks and prevents catastrophic forgetting. \emph{Code is available at} \url{https://github.com/NMS05/Adapter-Incremental-Continual-Learning-AST}.

\end{abstract}
\vspace{0.5em}

\begin{figure}[t]
\centering
\includegraphics[width=0.7\linewidth]{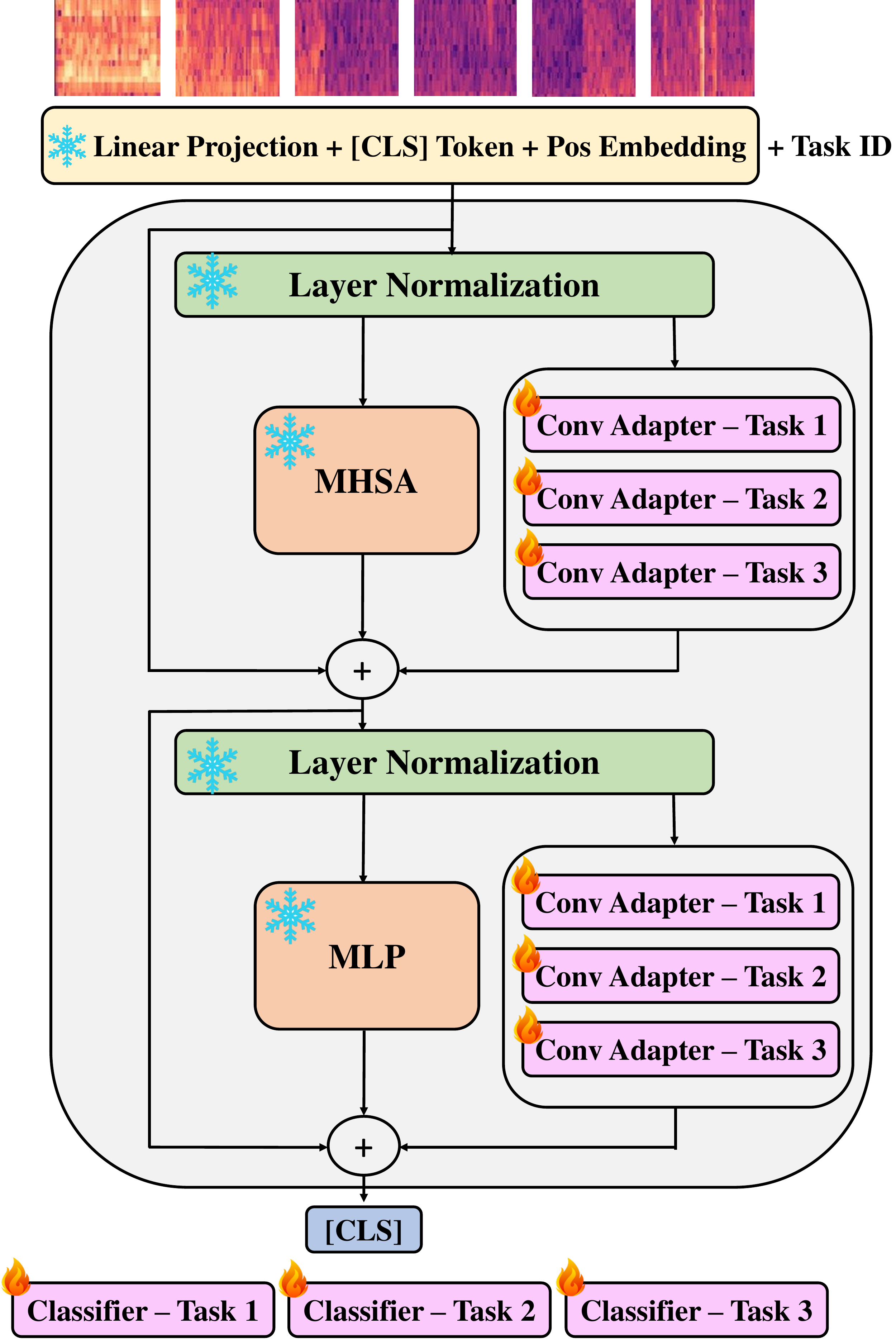}
\vspace{-0.5em}
\caption{Adapter Incremental Continual Learning of Audio Spectrogram Transformers.}
\label{fig:TICL}
\vspace{-2.0em}
\end{figure}

\section{Introduction}
% Continual Learning (CL) of new knowledge and acquisition of new skills are the desired characteristics of any intelligent machine. However, CL still remains a major challenge for Deep Learning. When neural networks are trained on a new task, they forget the knowledge gained in the previous task, a phenomenon known as “catastrophic-forgetting” [cite]. This is because the weights of the neural nets are optimized (new knowledge) for the new task only without any regard to the knowledge gained in the previous tasks. While catastrophic forgetting can be minimized by constraining the weights of neural nets [cite EMC, SI] or using data/pseudo-data of previous tasks [cite], a rather simple approach would be to use sub-networks within a large neural network [cite]. During training, only the sub-network is optimized for the new task while other parameters can be shared across tasks. Such a setup is called Task Incremental Continual Learning (TI-CL), which requires a “Task-ID” to route the data to the corresponding sub-network. As the model is trained on new tasks over time, the model size grows sub-linearly. In this paper, we explore task incremental continual learning of audio spectrogram classifiers with the recently introduced Audio Spectrogram Transformers (AST) [cite], which achieved state-of-the-art in several audio benchmarks. However, formulating AST in a TICL setup faces two major challenges – parameter and compute inefficiency.

Continual learning~\cite{de2021continual} of new knowledge and skill acquisition are the desirable traits for intelligent machines. However, in Deep Learning, neural networks may forget previous knowledge~\cite{french1999catastrophic} due to the optimization of network weights for new tasks, leading to catastrophic forgetting. Many works have been proposed to address this issue by constraining the weights of neural nets~\cite{kirkpatrick2017overcoming,zenke2017continual} or using data (pseudo-data) of previous tasks~\cite{li2017learning}. 
% Another approach called Task Incremental Continual Learning (TI-CL) can also 
A simple way to mitigate this issue is to assign task-specific sub-networks, where only a sub-network is optimized for new tasks while other parameters are task-independent and can be shared across tasks. This approach is particularly effective for Task Incremental Continual Learning (TI-CL), which requires a task-ID to route the data to the corresponding sub-network. As the model is incrementally trained on new tasks, its size grows sub-linearly.

This paper explores TI-CL of audio classifiers with Audio Spectrogram Transformers (AST)~\cite{gong2021ast}, which achieved state-of-the-art results on several audio benchmarks~\cite{piczak2015esc,warden2018speech,gemmeke2017audio}. However, there are two main issues with AST that must be addressed for sequential training: parameter inefficiency and computational inefficiency.

% \noindent\textbf{Parameter Inefficiency} – full fine-tuning was the de-facto standard for transfer learning with pretrained models. But transformer based models like AST have far more trainable parameters and this issue is more pronounced with ever increasing size of transformers from a few hundred million parameters to a few billion parameters. Also, transformers require more data because training such large number of parameters with very little data can lead to over-fitting.

\noindent\textbf{Parameter Inefficiency.} In TI-CL, the use of pre-trained transformer-based models like AST can lead to parameter inefficiency due to a large number of trainable parameters in full-finetuning for sequential tasks. This can cause overfitting, especially when the sequential tasks have limited data.

% \noindent\textbf{Compute Inefficiency} – the self-attention mechanism of transformers has quadratic computational complexity i.e. the number of computations increase quadratically with an increase in the number of tokens. Unlike RGB images, which have a fixed resolution and can also be resized, audio spectrograms cannot be resized. Hence, a long duration audio with a higher frequency resolution (Mel bins) results in larger spectrogram images, which drastically increases both the number of tokens and the number of computations.

\noindent\textbf{Computational Inefficiency.} The transformer's self-attention mechanism~\cite{vaswani2017attention} has quadratic computational complexity.
% The number of computations grow exponentially as the number of tokens increases. 
Hence, a large number of tokens extracted from larger spectrograms (from long duration audio) rapidly increases the number of computations.
However, audio spectrograms cannot be resized since their characteristics are determined by the audio duration and the number of frequency bins. Resizing audio spectrograms can lead to a loss of critical information and adversely affect their quality. Hence, transformer-based AST shows significant computational inefficiency when processing long-duration audio.

\vspace{-0.1em}

Therefore, we propose a TI-CL method based on AST and address the issues of parameter and computational efficiency. 
% Leveraging PET learning methods, we introduce Convolutional Adapters for AST to improve parameter efficiency.
We leverage Parameter Efficient Transfer (PET) methods to improve the parameter efficiency of AST. Our study evaluates the efficacy of various PET methods for AST on ESC-50~\cite{piczak2015esc} and SpeechCommandsV2~\cite{warden2018speech} benchmarks and proposes Convolutional Adapters to address parameter inefficiency. Note that the performance of PET methods for AST audio classifiers has not been studied before. The convolutional adapters perform as well as fully fine-tuned models in high-resource settings and even outperform them in low-resource settings with $<$5\% of the trainable parameters. 

\vspace{-0.1em}

\begin{figure}[t]
\centering
\includegraphics[width=0.85\linewidth]{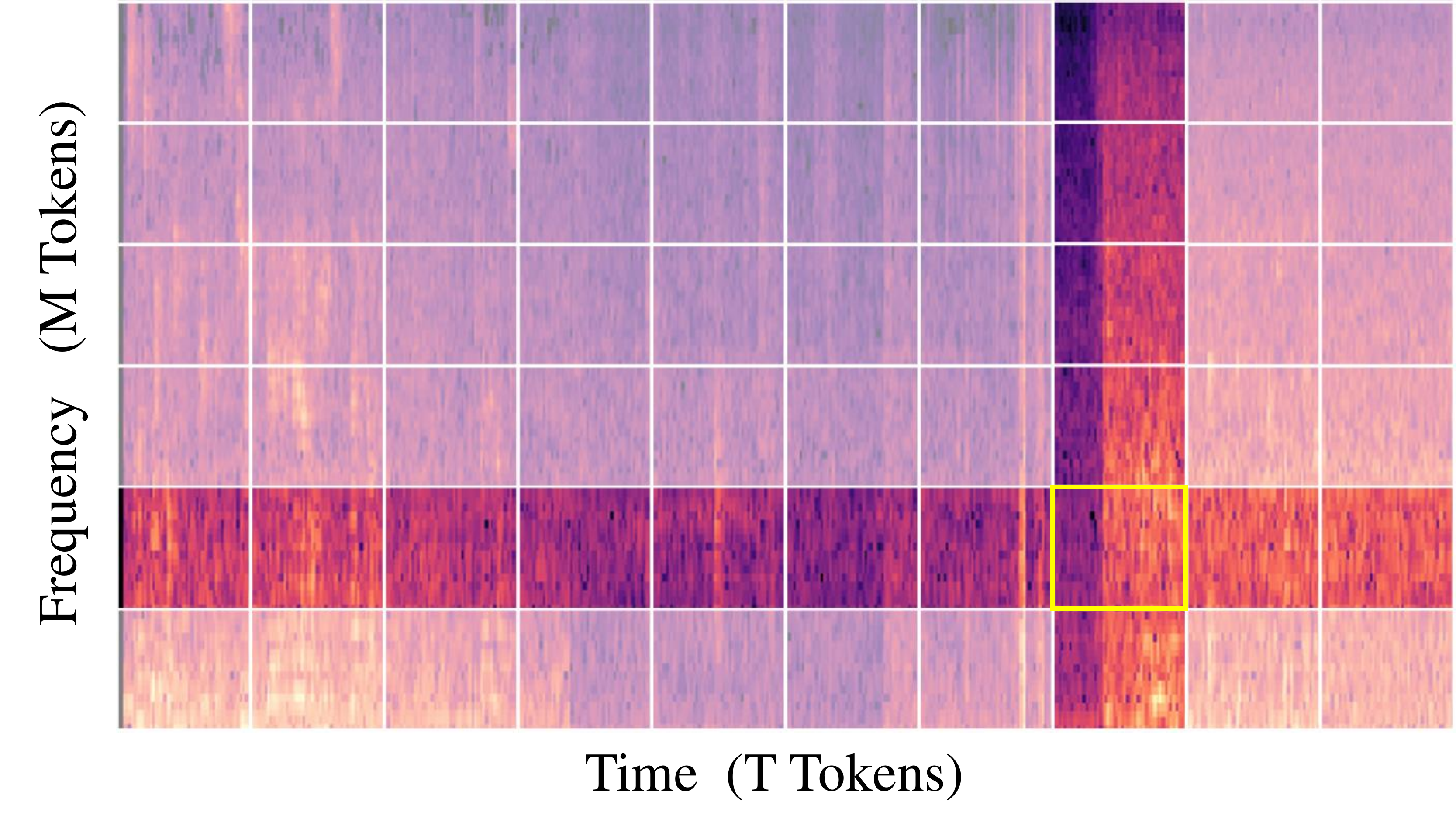}
\vspace{-1.0em}
\caption{Frequency-Time factorized Attention for a (yellow) token along the frequency and time axis.}
\label{fig:FTA}
\vspace{-2.0em}
\end{figure}

Next, we propose Frequency-Time factorized Attention (FTA) to address computational inefficiency in self-attention for long-duration audio spectrograms. Unlike traditional self-attention, FTA enables an arbitrary token to attend only to the frequency and temporal tokens that share the same position index in either axis, thereby leveraging the orthogonal nature of frequency and time in spectrograms (see Fig.~\ref{fig:FTA}). This factorization greatly reduces complexity and improves computational efficiency. % Our experimental results demonstrate that FTA achieves competitive performance with global self-attention while requiring only a fraction of the computations.
To achieve both parameter and computational efficiency, we combine Convolutional Adapter and FTA for TI-CL of audio classification. 
\vspace{-0.2em}

The main contributions of this paper can be summarized as follows,
\begin{itemize}
    \item We provide an empirical study on the performance of various PET methods for AST. 
    \vspace{-0.1em}
    \item We propose TI-CL of audio classifiers with parameter-efficient AST, using Convolutional Adapters.
    \item We introduce a novel Frequency-Time factorized Attention (FTA) for compute-efficient AST.
    \vspace{-0.1em}
   \item Through comprehensive experiments we demonstrate the advantages of the proposed approach for TI-CL of audio classifiers.
\end{itemize}

\vspace{-1.0em}
\section{Related work}

\subsection{Continual Learning for Audio}

To prevent catastrophic forgetting in continual learning, various methods have been proposed. For example, GIM~\cite{cossu2020continual} incrementally adds new modules to capture drifts in input distribution, DFWF~\cite{ma21b_interspeech} uses a knowledge distillation loss to preserve memory from the original model, and static memory networks~\cite{karam2022task} introduce static memory to reduce memory usage and model complexity. Few-shot CL~\cite{wang2021few} enables fast and interactive model updates in a few-shot learning framework to expand the audio classifier to recognize novel classes, while CTR~\cite{ke2021achieving} addresses both catastrophic forgetting and knowledge transfer issues with a pair of continual learning plugin modules.

\subsection{Parameter Efficient Transfer}

Many recent works have focused on efficient transfer learning and fine-tuning techniques for downstream tasks, such as Adapter for NLP~\cite{houlsby2019parameter} and similar methods like LoRA~\cite{hulora}, AdaptFormer~\cite{chenadaptformer}, and ConvPass~\cite{jie2022convolutional}. These methods achieve efficient fine-tuning by inserting small trainable bottleneck modules at different locations inside a transformer encoder while freezing other parameters during training. Commonly used methods involve a down projection followed by an up projection. Other methods tune specific parameters in the network, such as BitFit~\cite{zaken2021bitfit}, which adapts the model for different tasks by tuning the bias terms of the transformer layers, LayerNorm Tune, which tunes the affine transformation parameters in the encoder normalization layers, and Prompt Tuning~\cite{jia2022visual}, which optimizes a set of learnable latent tokens that are prepended to the input sequence at every encoder layer for transfer learning.

\begin{figure*}[t]
\centering
\includegraphics[width=0.85\linewidth]{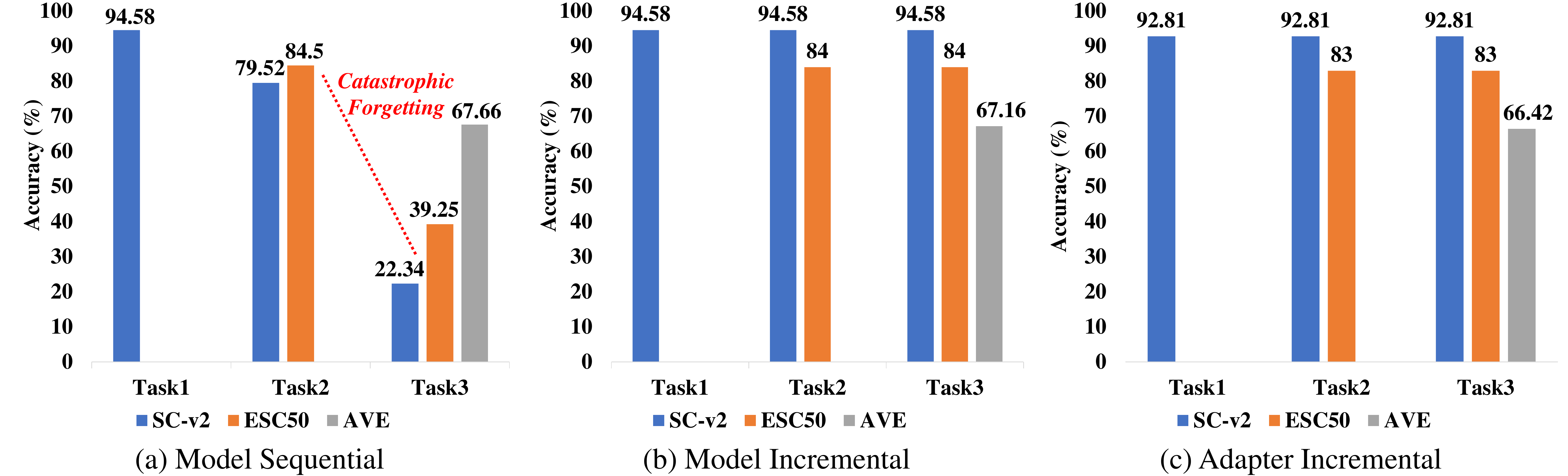}
\caption{Performance of the AST model in TI-CL setup for three training modes.}
\label{fig:main}
\vspace{-1.0em}
\end{figure*}

\vspace{-0.5em}
\section{Methodology}
% this part can be removed if the space is limited.
% We introduce the TI-CL setup and briefly recap the AST audio classifier in Sec 3.1. We then describe the continual learning of parameter-efficient adapter based AST in Sec 3.2 and finally explain the proposed compute-efficient FTA method in Sec 3.3.

\subsection{Continual Learning (CL) and AST audio classifier}

The objective of continual learning is to sequentially train a parameterized model $f_{\boldsymbol{\theta}}$ over a set of $n$ tasks $D \in \{D_1, D_2, ..., D_n\}$. Each task is defined by $D_i=(X_i, Y_i), i \in [1, n]$, where $X$ is a set of input samples and $Y$ is a set of corresponding labels. The parameterized function $f_{\boldsymbol{\theta}}: x \xrightarrow{} y$ maps the input $x \in X$ to the corresponding  label $y \in Y$ and the goal of CL is to train $f_{\boldsymbol{\theta}}$ such that it can correctly predict the label $y$ for an unseen arbitrary input $x$ sampled across $D$. 

If $D$ is an audio classification task, then $f_{\boldsymbol{\theta}}$ is a pre-trained AST model with total weights $\boldsymbol{\theta}$, $x \in X$ is a spectrogram image and $y \in Y$ is the corresponding audio class label. $f_{\boldsymbol{\theta}}$ extracts tokens $\boldsymbol{Z}=\{z_1,z_2,...,z_{MT+1}\}$ from $x$, where $z \in \mathbb{R}^d$, $M$ and $T$ denote the number of tokens in frequency and time axis, $d$ is the embedding dimension and $1$ denotes the class token.
% $f_{\boldsymbol{\theta}}$ extracts tokens of shape $\boldsymbol{Z} \in \mathbb{R} ^ {(MT+1) \times d}$ from $x$, where $M$ and $T$ denote the number of tokens in frequency and time axis, $d$ is the embedding dimension and $1$ denotes the class token.Interpolated position embeddings of the same shape are added to the input tokens. 
These tokens are processed by a series of 12 transformer encoders with Multi-Head Self-Attention (MHSA), Multi Layer Perceptron (MLP) and Layer Normalization (LN) sublayers, and can be formulated as,
% \vspace{-0.5em}
% The MHSA layer extracts a set of query $Q$ and key-value pairs $K,V$ with weights $W_Q,W_K,W_V \in \mathbb{R} ^ {d \times d}$ and the self-attention is computed as follows,
% \begin{equation}\small
% \label{eq1}
%     MHSA = Softmax \left(\cfrac{Q.K^{\intercal}}{\sqrt{d}}\right)V
% \end{equation}
\begin{equation}\footnotesize
\label{eq2}
\begin{split}
    \boldsymbol{Z'_l} &= MHSA(LN_1(\boldsymbol{Z_{l-1}})) + \boldsymbol{Z_{l-1}}, \\
    \boldsymbol{Z_l} &= MLP(LN_2(\boldsymbol{Z'_l})) + \boldsymbol{Z'_l},
\end{split}   
\end{equation}
% \begin{equation}\small
% \label{eq3}
%     \boldsymbol{Z_l} = MLP(LN_2(\boldsymbol{Z'_l})) + \boldsymbol{Z'_l},
% \end{equation}
\noindent where $l$ denotes the layer number and $\boldsymbol{Z_l}$ is the extracted tokens from layer $l$.

\subsection{Adapter Incremental Continual Learning of AST}
% this part should introduce parameter efficient 
% While full fine-tuning optimizes $\theta$ directly, Adapters insert a set of additional learnable parameters while keeping $\theta$ frozen. 
% 
Task Incremental Continual Learning is one of the three scenarios for CL~\cite{van2019three}, where it assumes that the tasks $D_i$ are disjoint and the task ID $i$ is known both during training and inference. Full-finetuning $f_{\boldsymbol{\theta}}$ on the sequential tasks by optimizing $\boldsymbol{\theta}$ may not be efficient and may lead to the overfitting issue. A parameter incremental approach to solve TI-CL involves training a parameterized network with multiple task-specific sub-modules denoted as $f_{\boldsymbol{\theta}+\delta{\boldsymbol{\theta}}}$, where  $\theta$ is the shared task-independent parameter, $\delta{\boldsymbol{\theta}} \in \{\boldsymbol{\theta_1}, \boldsymbol{\theta_2}, ..., \boldsymbol{\theta_n}\}$ are the task-specific parameters and $\boldsymbol{\theta}$ is much larger than $\delta\boldsymbol{\theta}$.

We propose an adapter incremental method for TI-CL called Adapter Incremental Continual Learning (AI-CL), where a Convolutional Adapter (CA) is incrementally added and trained for each task while keeping the shared $\boldsymbol{\theta}$ frozen. We denote the weights of task-specific CA as $\delta\boldsymbol{\theta_i}$ for every new task $D_i$. CA has a bottleneck structure, which consists of a down-projection followed by an up-projection with an additional 2D convolution layer in between. The inputs tokens are reshaped to $M \times T$ before the convolution operation, with the exception of the class token, and then reverted back to its original shape before up-projection. CA processes arbitrary length input tokens $\boldsymbol{z} \in \mathbb{R}^d$ as,
\begin{equation}\footnotesize
\label{eq3}
    CA(\boldsymbol{z}) = \boldsymbol{W_{up}}(GELU(Conv2D(\boldsymbol{W_{down}}(\boldsymbol{z})))),
\end{equation}
\noindent where $\boldsymbol{W_{down}} \in \mathbb{R} ^ {d \times d'}$, $\boldsymbol{W_{up}} \in \mathbb{R} ^ {d' \times d}$ and $d' << d$. 
% $\boldsymbol{W_{down}}$ and $Conv2D$ is followed by GELU non-linear activation. 
CA runs parallel to both MHSA and MLP layers, which can be represented as,
\vspace{-0.45em}

% \begin{equation}\footnotesize
% \label{eq4}
\begin{align*}
    \boldsymbol{Z'_l} &= MHSA(LN_1( \boldsymbol{Z_{l-1}})) +  \boldsymbol{Z_{l-1}} + CA_1(LN_1( \boldsymbol{Z_{l-1}})), \\
    \boldsymbol{Z_l} &= MLP(LN_2(\boldsymbol{Z'_l})) + \boldsymbol{Z'_l} + CA_2(LN_2(\boldsymbol{Z'_l})).
\end{align*}
% \end{equation}

% \begin{equation}\footnotesize
% \label{eq5}
%     \boldsymbol{Z_l} = MLP(LN_2(\boldsymbol{Z'_l})) + \boldsymbol{Z'_l} + CA_2(LN_2(\boldsymbol{Z'_l})).
% \end{equation}

\vspace{-0.45em}
\noindent The proposed AI-CL method using CA is parameter efficient since only the CA weights $\delta\boldsymbol{\theta_i}$ are trainable and saving these weights also occupies less storage. The backbone weights $\boldsymbol{\theta}$ are frozen and shared across tasks, both during the training and inference stage. During inference, when a test audio spectrogram $x$ is passed along with the task ID $i$, the AST model routes the tokens $\boldsymbol{Z}$ to the corresponding CA with the parameter $\delta\boldsymbol{\theta_i}$ and the corresponding classifier. The AST model with multiple task-specific CAs is illustrated in Fig \ref{fig:TICL}.

\vspace{-0.5em}

\subsection{Frequency-Time factorized Attention (FTA)}
% need to keep the same notation format as previous ones and present in math as well.
% While the parameter-efficient Adapter Incremental method using CA is effective for AST, 
While the AI-CL approach is parameter-efficient, the use of self-attention in AST results in a quadratic increase in computations (\emph{i.e.}, the number of floating point operations or FLOPS) for larger spectrograms. To address this issue, prior alternatives to self-attention either limit self-attention to a local window~\cite{liu2021swin} or factorize self-attention along two orthogonal axis~\cite{arnab2021vivit}, but they were developed for images and videos. 

Inspired by the factorization approach~\cite{arnab2021vivit,tan2022pure}, we propose Frequency-Time factorized Attention (FTA) in the AI-CL method as shown in Figure \ref{fig:FTA}. It factorizes self-attention across the frequency and time axis of a spectrogram, by masking out the undesired tokens. This approach makes AST more computationally efficient, with attention along the frequency (vertical) axis learning the distribution of various frequency components at a given time interval, and attention along the time (horizontal) axis learning how a frequency component evolves over time. The only exception is the $[CLS]$ token, which attends to all the tokens (including itself) since it must summarize the semantic information in a spectrogram. For a token $\boldsymbol{Z} \in \mathbb{R} ^ {(MT+1) \times d}$, the computation complexity $\mathcal{O}$ of Global Self-Attention (GSA) and FTA can be calculated as follows,
% \vspace{-0.2em}
\begin{equation}\footnotesize
\label{eq5}
\begin{split}
     \mathcal{O}_{GSA} &= (MT+1)^2*d, \\
    \mathcal{O}_{FTA} &= (MT(M+T+1)+1)*d,
\end{split}
% \vspace{-0.5em}
\end{equation}
% \begin{equation}\small
% \label{eq6}
%     FTA = MT(M+T+1)+1.
% \end{equation}
where $(M+T)<<MT$. Thus, when $M$ and $T$ grow, FTA has much fewer computations than GSA. Empirically, we show that the proposed Frequency Time factorized Attention (FTA) achieves competitive performance to global self-attention with only a fraction of the computations.

\begin{table*}[t]
\caption{Computational efficiency of the proposed FTA. $k$ denotes the factor of GSA computations required by FTA.}
\vspace{-0.5em}
\label{tab:FTA advantages}
\centering
\resizebox{0.88\textwidth}{!}{%
\begin{tabular}{ccccccccc}
\toprule
Dataset & Duration   & Spectrogram Shape & Freq (M Tokens) & Time (T Tokens) & $\mathcal{O}_{GSA}/d$    & $\mathcal{O}_{FTA}/d$   & $k$     \\ 
\midrule
SCv2    & 1s  & [128,101]         & 12              & 9               & 11881   & 2377   & 0.2   \\
ESC-50  & 5s  & [128,501]         & 12              & 49              & 346921  & 36457  & 0.105 \\
AVE     & 10s & [128,1006]        & 12              & 100             & 1442401 & 135601 & 0.094 \\
\bottomrule
\end{tabular}}
\vspace{-1.0em}
\end{table*}

\section{Results}
% Again, here if not much space can remove.
% We describe the experimental setup in Sec 4.1 and evaluate various PET methods in Sec 4.2. We then demonstrate the advantages of the proposed TI-CL setup in Sec 4.3 and finally compare global self-attention with the proposed FTA in Sec 4.4

\vspace{-0.5em}

\subsection{Experimental Setup}
% the introduction of the dataset can be more concise.
% \textbf{Datasets} - The following datasets were used for PET evaluation and TI-CL experiments.
% \begin{itemize}
%     \item ESC-50~\cite{piczak2015esc} dataset consists of 2,000 5-second environmental audio recordings organized into 50 classes. We follow the standard 5-fold cross-validation unless otherwise specified.
%     \item The Speech Commands V2 (SCv2)~\cite{warden2018speech} dataset consists of 105k 1-second recordings of 35 speech classes. We follow the standard training and test set split with 84,843 and 11,005 samples respectively.
%     \item AVE~\cite{tian2018audio} is an event localization dataset of 4143 samples covering 28 audio events with a duration of 10 seconds (long duration). We only use the audio modality and follow the original train-test split for audio classification.
% \end{itemize}

% revised version:
\textbf{Datasets.} The datasets used for PET evaluation and TI-CL experiments are:
\begin{itemize}
    \item ESC-50~\cite{piczak2015esc}, which contains 2,000 5-second audio recordings organized into 50 classes for environmental sound classification. The standard 5-fold cross-validation is used unless otherwise specified.
    \item Speech Commands V2 (SCv2)~\cite{warden2018speech}, which includes 105k 1-second recordings of 35 speech classes for speech recognition. The standard training and test set split is used with 84,843 and 11,005 samples respectively.
    \item AVE~\cite{tian2018audio}, an event localization dataset of 4,143 samples covering 28 events with a duration of 10 seconds (long duration). Only the audio modality is used, and the original train-test split for audio classification is followed.
\end{itemize}

% \noindent\textbf{Model} - The AST model is an ImageNet pre-trained ViT/B-16 model with 12 transformer encoders. The input audio waveform is converted into a log mel spectrogram with 128 Mel bins, 25ms Hamming window, and a hop length of 10ms, without any data augmentation. Tokens are extracted using a convolutional feature extractor with $kernel\_size=16, stride=10$ and $d=768$ with position embeddings added via bilinear interpolation. The model is trained using Adam optimizer (learning rate = $3e-4$) with cross-entropy loss on \{ESC-50, SCv2, AVE\} datasets with a batch size of \{32, 128, 12\} for epochs \{20, 5, 15\} respectively.

% revised version:
\noindent\textbf{Model.} Our system is built upon the AST model, a ViT/B-16 model with 12 transformer encoders pre-trained on the ImageNet-21k dataset (weights obtained from timm library). We process audio input by converting the waveform into a log mel spectrogram with 128 Mel bins, a 25ms Hamming window, and a hop length of 10ms, without any data augmentation. Tokens are extracted using a convolutional feature extractor with a kernel size of 16, a stride of 10, and a dimensionality of 768, with position embeddings added via bilinear interpolation. The model is trained using Adam optimizer with a learning rate of 3e-4 and cross-entropy loss, with batch sizes of 128/32/12 for the SCv2/ESC-50/AVE datasets. We train the model for 5/20/15 epochs on the respective datasets.
%%%%%%%%%%%%%%%%%%%%%%%%%%%%%%%%%%%%%%%%%%%%%%%%%%%%

\begin{table}[t]
\caption{Evaluation of PET methods for AST.}
\vspace{-0.5em}
\label{tab:PET comparison}
\centering
\resizebox{0.45\textwidth}{!}{%
\begin{tabular}{cccc}
% \hline
\toprule
\multirow{2}{*}{\textbf{Method}} & \multirow{2}{*}{\textbf{Params (Million)}} & \multicolumn{2}{c}{\textbf{Accuracy (\%)}} \\ \cline{3-4} & \\[-1.0ex]
                 &                & \textbf{ESC-50} & \textbf{SCv2} \\ %\hline
                 \midrule
Linear           & 0.26           & 71.05          & 81.44                      \\
LayerNorm Tune   & 0.27           & 72.75          & 89.2                       \\
BitFit~\cite{zaken2021bitfit}           & 0.32           & 72             & 87.91                      \\
AdaptFormer~\cite{chenadaptformer}      & 1.43           & 83             & 92.3                       \\
Prompt Tuning~\cite{jia2022visual}          & 2.17           & 78.85          & 91.64                      \\
LoRA~\cite{hulora}             & 2.6            & 79.05          & 92.14                      \\
Houlsby~\cite{houlsby2019parameter}          & 2.62           & 69.75          & 90.83                      \\
\rowcolor{Gainsboro!60}ConvPass~\cite{jie2022convolutional}         & \textbf{3.5}   & \textbf{83.3}  & \textbf{93.42}             \\ \hline
Full Fine Tuning & \textbf{86.33} & \textbf{82.3}  & \textbf{94.58}             \\ %\hline
\bottomrule
\end{tabular}
}
\vspace{-0.5em}
\end{table}
% the alignment between each lines are not very consistent and the font size is small.
%%%%%%%%%%%%%%%%%%%%%%%%%%%%%%%%%%%%%%%%%%%%%%%%%%%%%%%%%%%%%

\vspace{-0.5em}
\subsection{Evaluation of PET methods}
\vspace{-0.5em}
% It is important to state the motivation for doing such a comparison and how it will affect this work/study.
% We evaluate various PET methods (discussed in Sec 2.2) on ESC-50 and Speech Commands V2 datasets, and the results are presented in Table \ref{tab:PET comparison}. Note that for PET evaluation we use global self-attention and not the proposed FTA. It can be seen that adapters outperform other PET methods when they are in a parallel configuration [cite convpass and adaptformer]. The convolutional adapter performs on-par with full fine-tuning in a high resource setting (2.7k samples per class) like Speech Commands V2 and even surpasses them in a low resource setting (40 samples per class) like ESC-50 with less than 5\% of the trainable parameters. Through ablation studies we also find that the performance of ConvPass scales well with increase in parameters when compared to AdaptFormer, thus making it our primal choice for TI-CL formulation.

% Here is a revision:
While several PET methods have been proposed for NLP and Vision tasks, their effectiveness in audio classification remains largely unexplored. In this study, we evaluated several PET methods on the ESC-50 and SCv2 datasets, and found that AdaptFormer~\cite{chenadaptformer} and ConvPass~\cite{jie2022convolutional} achieved the highest performance (see Table \ref{tab:PET comparison}). The Linear method was simply adding a trainable linear layer for classification in Table~\ref{tab:PET comparison}. Notably, ConvPass achieved comparable performance to full fine-tuning on SCv2 (with 2.7k samples per class), and even outperformed it on ESC-50 (with only 40 samples per class) while using less than 5\% of trainable parameters. 
% Our results suggest that adapters like ConvPass and AdaptFormer which run parallel to MHSA/MLP layers can achieve better performance than Houlsby, which is in a sequential configuration. 
% This was emphasized more in AdaptFormer paper!!!
The evaluation provides compelling evidence for the effectiveness of a parameter efficient strategy. Therefore, we adopted the Convolutional Adapter for further investigation in TI-CL.

%%%%%%%%%%%%%%%%%%%%%%%%%%%%%%%%%%%%%%%%%%%%%%%%%%%
\begin{table}[t]
\caption{Comparison of parameter and storage cost for three training modes in TI-CL setup.}
\vspace{-0.5em}
\label{tab:TICL params}
\centering
\resizebox{0.45\textwidth}{!}{%
\begin{tabular}{ccccc}
\toprule
       & Trainable Params & Total Params & Storage \\
\midrule
Model Seq. & 86.5M            & 86.62M       & 348MB      \\
Model Inc. & 86.5M            & 259.63M      & 1.02GB    \\
\rowcolor{Gainsboro!60}Adapter Inc. & 3.5M                & 96.6M            & 47MB  \\ 
                    \bottomrule
\end{tabular}
}
\vspace{-0.5em}
\end{table}

% \begin{table}[t]
% \caption{Number of trainable parameters on the sequence of tasks. The total parameters are shown in the brackets.}
% \label{tab:TICL params}
% \centering
% \resizebox{0.45\textwidth}{!}{%
% \begin{tabular}{cccc}
% \toprule
%         & Task1  & Task2  & Task3  \\ 
% \midrule
% Model Sequential & 86.53 (86.56)  & 86.53 (86.6)  & 86.53 (86.62)  \\ 
%  \hline
% Model Incremental &  86.13 (86.13) & 86.5 (172.63)   & 87 (259.63)  \\
%  \hline
% Adapter Incremental &\rowcolor{Gainsboro!60} 1.448 (87.568)  & 1.459 (89.027)  & 1.443 (90.47)  \\ 
% \bottomrule
% \end{tabular}}
% \end{table}

% \begin{table}[t]
% \caption{Merits of Adapter based TI-CL}
% \label{tab:AICL advantages}
% \centering
% \resizebox{0.45\textwidth}{!}{%
% \begin{tabular}{cccc}
% \toprule
%                     & Storage             & Compute (Trainable Params) & Performance \\
% \midrule
% Model Sequential    & High (340MB)        & High (86.5M)               & Degrades    \\
% Model Incremental   & Very High   (933MB) & High (86.5M)               & Stable      \\
% \rowcolor{Gainsboro!60}Adapter Incremental & Low (16MB)          & Low (1.45M)                & Stable      \\
% \bottomrule
% \end{tabular}
% }
% \end{table}

%%%%%%%%%%%%%%%%%%%%%%%%%%%%%%%%%%%%%%%%%%%%%%%%%%%%%%%%%%%%%
\begin{table}[t]
\caption{Performance of FTA vs GSA on three tasks.}
\vspace{-0.5em}
\label{tab:FTA performance}
\centering
\resizebox{0.33\textwidth}{!}{%
\begin{tabular}{cccc}
\toprule
\multirow{2}{*}{Method} & \multicolumn{3}{c}{Accuracy   (\%)}      \\ \cline{2-4} & \\[-1.5ex]
                        & SCv2 & ESC-50 & AVE (Audio) \\ 
\midrule
GSA          & 93.57            & 85.25  & 69.1        \\
\rowcolor{Gainsboro!60} FTA                     & 92.81            & 83     & 66.42       \\
\bottomrule
\end{tabular}
}
\vspace{-0.8em}
\end{table}

%%%%%%%%%%%%%%%%%%%%%%%%%%%%%%%%%%%%%%%%%%%%%%%%%%%%%%%%%%%%%
\vspace{-0.5em}
\subsection{Adapter Incremental Continual Learning of AST}

% The training mode and formulation can be better combined. The formulation can be shortened to the point that each time, the model will be trained on one specific dataset and the previous task is no longer available for training. (task 1:xxx dataset, task 2: xxx dataset, and task 3, xxx dataset)
% \textbf{Formulation} - The AST model is trained on three tasks (SCv2, ESC-50 and AVE) in a sequential order as follows,
% \begin{itemize}
%     \item Task 1 - Only SCv2 data is available and the model is evaluated on this dataset.
%     \item Task 2 - It is the next time step and only ESC-50 data is available. The SCv2 training data is no longer available and the model is evaluated on ESC-50 and SCv2 test set.
%     \item Task 3 - Only AVE data is available in this (next) time step and the training data of SCv2 and ESC-50 is no longer available. The model is evaluated on AVE, ESC-50 and SCv2 test sets.
% \end{itemize}

% revised version:
\textbf{Formulation.} The TI-CL setup consists of three tasks: SCv2, ESC-50, and AVE, which are performed in a sequential order. In each task of the TI-CL, only the corresponding dataset is available for training and the datasets from previous tasks are no longer available. Only the test data of previous tasks are used to evaluate the model performance after training on current task.

% \noindent\textbf{Training Modes} -  To demonstrate the advantages of the proposed approach  the AST model is trained in three different modes, all of which follow the above sequential training order.
% \begin{enumerate}
%     \item Model Sequential - This is a naive setting where the same AST model is trained over and over again on new tasks.
%     \item Model Incremental - For every new task, a new AST model is trained and they are independent of each other.
%     \item Adapter Incremental - This is the proposed approach described in Sec 3, where new adapter modules are added to the frozen backbone for new tasks.
% \end{enumerate}

% revised version:
\noindent\textbf{Training Modes.} To demonstrate the proposed approach's advantages, we trained the AST model in three different modes, following the sequential training order. These modes are:
\begin{itemize}
    \item Model Sequential: The same AST model is trained repeatedly on new tasks.
    \item Model Incremental: For every new task, a new AST model is trained independently.
    \item Adapter Incremental: The proposed approach described in Section 3, where new adapter modules are added to the frozen backbone with FTA for new tasks.
\end{itemize}

The first two modes rely on GSA and the ESC-50 task is evaluated with single fold.

% Analysis of the Convolutional Adapter, how it performs, and reveal the reason behind it if possible.
% \noindent\textbf{Performance vs Parameter-Efficieny} - The performance of the AST model for three training modes is shown in Figure \ref{fig:main} and the evolution of model parameters over time is presented in Table \ref{tab:TICL params}. The phenomenon of catastrophic forgetting is observed in Model Sequential setting where the model weights optimized for a new task forget the knowledge gained from previous tasks. Model Incremental setting resolves this issue because the models are independent of each other, but from Table \ref{tab:TICL params}, the total number of parameters explode over time. Moreover, both these training modes require 20x trainable parameters. The proposed adapter based TI-CL formulation combines the best of both worlds by delivering stable performance (without catastrophic forgetting), yet with minimal trainable parameters and a summary of its merits is presented in Table \ref{tab:AICL advantages}.

% revised version:
\noindent\textbf{Performance vs Parameter-Efficieny.} Figure \ref{fig:main} displays the performance of the AST model for three training modes. In the Model Sequential setting, catastrophic forgetting occurred, where the model weights optimized for a new task forgot the knowledge gained from previous tasks, leading to a significant performance drop. However, Model Incremental setting trained the models independently for each task, thereby resolving this issue. The proposed Adapter Incremental method also addressed the catastrophic forgetting issue by training independent task-specific adapter modules and showed competitive performance on all three tasks. However, the Model Sequential and Model Incremental settings were less efficient than the Adapter Incremental method in terms of total model parameters and trainable parameters, as illustrated in Table \ref{tab:TICL params}. Note that the total number of parameters were those required for inference upon the completion of sequential training on three tasks. The Model Incremental setting had a large number of total parameters, and both Model Sequential and Model Incremental settings required nearly 25 times more trainable parameters than Adapter Incremental. Overall, the proposed Adapter Incremental method for TI-CL combined the best of performance and parameter efficiency, delivering stable performance and minimizing the number of trainable parameters. Also, Adapter Incremental setting has substantially lower storage cost because only the adapter weights need to be saved, unlike the other two setting which stores the weights of the entire models(s).
% Table \ref{tab:AICL advantages} summarizes its merits and storage refers to the fully trained model (or module) weights saved as tensors.

% \subsection{Impact of FTA}
% The advantages of the proposed Frequency-Time factorized Attention over global self-attention in terms of compute (FLOPS) for the three datasets is presented in Table \ref{tab:FTA advantages}. It can be seen that FTA requires only a fraction of the computations required by self attention and this is more evident with larger (long audio duration) spectrograms. For the adapter based AST model, the performance of FTA vs global self-attention for the three datasets are given in Table \ref{tab:FTA performance}. FTA performs competitively with global self-attention, yet with only a fraction of the computations.

\subsection{Impact of FTA}
We conducted a study to compare the computational efficiency and performance of our proposed FTA with Global Self-Attention (GSA) on three datasets, each with varying maximum audio durations. In Table \ref{tab:FTA advantages}, we summarize the details of our comparison. Our results showed that FTA required significantly fewer computations than GSA, especially with longer audio durations (the larger spectrograms). To further evaluate the performance of FTA and GSA, we implemented both methods using the Convolutional Adapter model and measured their audio classification accuracies on the three datasets. The results are presented in Table \ref{tab:FTA performance}. We found that FTA performed competitively with GSA in terms of accuracy, but with only a fraction of the computational resources required by self-attention. Overall, our study demonstrates that FTA is a promising approach for audio classification tasks, as it achieves comparable accuracy to GSA while using significantly fewer computational resources.

\section{Conclusions}

% In this work, we proposed a new method called Adapter Incremental Continual Leaning for TI-CL of AST audio classifiers. To improve the parameter efficiency, we introduced Convolutional Adapters for TI-CL. To enhance compute efficiency for long audios, we proposed Frequency-Time factorized Attention. Through experiments, we demonstrated that our proposed method is both parameter-efficient and compute-efficient, and enabled continual learning with minimal resources, which can e efficiently scaled for a large number of tasks.

% In this work, we proposed a new method called Adapter Incremental Continual Learning (AI-CL) for audio classification in the context of time-incremental continual learning (TI-CL) of automatic speech recognition (AST) systems. AI-CL improved parameter efficiency with the introduction of Convolutional Adapters for TI-CL. To enhance compute efficiency for longer audio streams, we proposed a new method called Frequency-Time factorized Attention. Our experiments have shown that AI-CL is both parameter-efficient and compute-efficient. AI-CL enables continual learning with minimal resources, which can be scaled effectively for a large number of tasks. 
In this work, we proposed a new method called Adapter Incremental Continual Learning (AI-CL) for audio classification in the context of Task Incremental Continual Learning (TI-CL) of AST audio classifiers. AI-CL improved parameter efficiency with the introduction of Convolutional Adapters for AST. To enhance compute efficiency for longer audio streams, we proposed a new method called Frequency-Time factorized Attention. Our experiments have shown that AI-CL is both parameter-efficient and compute-efficient. AI-CL enables continual learning with minimal resources, which can be scaled effectively for a large number of tasks.

\section{Acknowledgements}

This work was carried out at the Rapid-Rich Object Search (ROSE) Lab, Nanyang Technological University, Singapore. The research is supported by the DSO National Laboratories, under the project agreement No. DSOCL21238.

\bibliographystyle{IEEEtran}

\bibliography{mybib}

\end{document}